\documentclass[epj]{svjour}
\usepackage{graphics}
\newcommand{\be}{\begin{equation}}
\newcommand{\ee}{\end{equation}}
\newcommand{\bea}{\begin{eqnarray}}
\newcommand{\eea}{\end{eqnarray}}
\begin{document}
\title{On the coupling $g_{f_0K^+K^-}$  and the structure of $f_0(980)$}
\subtitle{}
\author{Fulvia De Fazio
}                     
\institute{Istituto Nazionale di Fisica Nucleare - Sezione di
Bari, Italy}
\date{Received: date / Revised version: date}
%
\abstract{ We use light-cone QCD sum rules to evaluate the strong
coupling $g_{f_0 K^+ K^-}$ which enters in several analyses
concerning the scalar $f_0(980)$ meson. The result is  $6.2\le
g_{f_0 K^+ K^-}\le 7.8$ GeV.
} 
\maketitle
\section{Introduction}
\label{intro} The nature of light scalar mesons  still needs to be
unambiguously established \cite{rassegna,Close:2002zu}. Their
identification is made problematic since both quark-antiquark
$(q{\bar q})$ and non $q{\bar q}$ scalar states are expected to
exist in the energy regime below 2 GeV. For example, lattice QCD
and QCD sum rule analyses indicate that the lowest lying glueball
is a $0^{++}$ state with mass in the range 1.5-1.7~GeV
\cite{Morningstar:1999rf}. Actually, the observed light scalar
states are too numerous to be accomodated in a single $q \bar q$
multiplet,  and therefore it has been suggested  that some of them
escape the quark model interpretation. Besides glueballs, other
interpretations include multiquark states and quark-gluon
admixtures.
\par
Particularly debated  is the nature of  $f_0(980)$. Among the
oldest suggestions, there is the proposal that  confinement could
be explained by the existence of a state with vacuum quantum
numbers and mass close to the proton mass \cite{Close:1993ti}. On
the other hand, following the quark model and considering the
strong coupling to kaons, $f_0(980)$ could be interpreted as an
$s{\overline s}$ state
\cite{tornqvist,Tornqvist:1995kr,roos,scadron}. However, this does
not explain the mass degeneracy between $f_0(980)$ and $a_0(980)$
interpreted as a $({u{\overline u} -d {\overline d})/\sqrt{2}}$
state. A four quark $qq{\overline{qq}}$ state interpretation has
also been proposed \cite{jaffe}. In this case, $f_0(980)$ could
either be nucleon-like \cite{ivan}, {\it i.e.} a bound state of
quarks  with symbolic quark structure $f_0={s{\overline s}({ u
{\overline u}+d {\overline d})/ \sqrt{2}}}$, the $a_0(980)$ being
$a_0=s {\overline s}( u {\overline u}-d {\overline d}) /
\sqrt{2}$, or deuteron-like, {\it i.e.} a bound state of hadrons.
If $f_0$ is a bound state of hadrons, it is usually referred to as
a $K {\overline K}$ molecule \cite{isgur,closebook,kaminski,shev}.
In the former of these two possibilities  mesons are treated as
point-like, while in the latter they should be viewed as extended
objects. The identification of the $f_0$ and of the other lightest
scalar mesons with the Higgs nonet of a hidden U(3) symmetry has
also been suggested \cite{Tornqvist:2002bx}. Finally, a different
interpretation consists in considering  $f_0(980)$  as the result
of a process in which strong interaction enriches a pure ${\bar
q}q$ state with other components, such as $|{K{\bar K}}\rangle$, a
process known as hadronic dressing
\cite{Tornqvist:1995kr,Boglione:1996uz}; such an interpretation is
supported in
\cite{Close:2002zu,tornqvist,Tornqvist:1995kr,scadron,Tornqvist:2000ju,Shakin:gn,DeFazio:2001uc}.
\par
The radiative $\phi \to f_0 \gamma$ decay mode has been identified
as  an effective tool to discriminate among the various scenarios
\cite{ivan,closebook,Close:2001ay}. As a matter of fact, if $f_0$
has a pure strangeness component $f_0=s{\bar s}$, the dominant
$\phi \to f_0 \gamma$ decay mechanism is the direct transition,
while  in the four-quark scenario $\phi \to f_0 \gamma$ is
expected to proceed through kaon loops with a branching fraction
depending on the specific bound state structure
\cite{closebook,Close:2001ay}.
\par
An important hadronic parameter is the strong coupling $g_{f_0 K^+
K^-}$. Indeed, the kaon loop diagrams contributing to $\phi \to
f_0 \gamma$ are expressed in terms of  $g_{f_0 K^+ K^-}$, as well
as in terms of  $g_{\phi K^+ K^-}$ which can be inferred from
experimental data on $\phi$ meson decays. In the present paper, we
report on a study  \cite{Colangelo:2003jz} devoted to determining
$g_{f_0 K^+ K^-}$ by light-cone QCD sum rules
\cite{lightconesr,Colangelo:2000dp}. Such an analysis is presented
in Section \ref{calculation}, while comparison with experimental
and  theoretical determinations is given in Section
\ref{conclusions}.
\section{ $g_{f_0 K^+ K^-}$ by light cone QCD Sum rules}
\label{calculation} In order to evaluate the strong coupling
$g_{f_0 K^+ K^-}$, defined by the matrix element: \be < K^+(q)
K^-(p)|f_0(p+q) >=g_{f_0 K^+ K^-}\,\,\, , \label{gf0kk} \ee we
consider the correlation function
\be T_\mu(p,q)=i \int d^4x \, e^{i p \cdot x} \,
\langle{K^+(q)}|T[J_\mu^K(x) J_{f_0}(0)]|0\rangle \,\,\, ,
\label{corr} \ee where $J_\mu^K={\bar u}\gamma_\mu \gamma_5 s$ and
 $J_{f_0}={\bar s}s$. The
external kaon state has four momentum $q$, with $q^2=M_K^2$. The
choice of the  $J_{f_0}={\bar s}s$ current does not imply that
$f_0(980)$ has a pure ${\bar s}s$ structure, but it simply amounts
to assume that  $J_{f_0}$ has a non-vanishing matrix element
between the vacuum and  $f_0$ \cite{DeFazio:2001uc,Aliev:2001mm}.
Such a matrix element, as mentioned below, has been derived by the
same sum rule method.

Exploiting Lorentz invariance, $T_\mu$ can be written in terms of
two independent invariant functions, $T_1$ and $T_2$: $ T_\mu(p,q)
= i \, T_1(p^2,(p+q)^2) \, p_\mu + T_2(p^2,(p+q)^2) \, q_\mu $.
 The  general strategy of QCD sum rules consists
in representing $T_\mu$ in terms of the contributions of hadrons
(one-particle states and the continuum) having non-vanishing
matrix elements with the vacuum and the currents ($J_\mu^K$ and
$J_{f_0}$ in the present case), and matching such a representation
with a QCD expression computed in a suitable region of the
external momenta $p$ and $p+q$ \cite{Craigie:1982ng}.

Let us consider, in particular,  the invariant function $T_1$
that can be represented  by a dispersive formula in the two
variables $p^2$ and $(p+q)^2$:
\be T_1(p^2,(p+q)^2)= \int ds ds^\prime {\rho^{had}(s,s^\prime)
\over (s-p^2) [s^\prime-(p+q)^2]} \, . \label{disp} \ee The
hadronic spectral density $\rho^{had}$ gets contribution from the
single-particle  states $K$ and $f_0$, for which we define
current-particle matrix elements: \be \langle f_0(p+q)|J_{f_0}|0
\rangle = M_{f_0} {\tilde f} \,\,\, , \,\,\, \langle 0 |J_\mu^K| K
(p)\rangle = i f_K p_\mu \,\,\, , \label{fk} \ee as well as from
higher resonances and a continuum of states that we assume to
contribute in a domain $D$ of the $s,s^\prime$ plane, starting
from two thresholds $s_0$ and $s_0^\prime$. Therefore, neglecting
the  $f_0$  width, the spectral function $\rho^{had}$ can be
modeled as: \bea \rho^{had}(s,s^\prime)&=& f_K M_{f_0} {\tilde f}
g_{f_0 K^+ K^-} \delta(s-M^2_K) \delta(s^\prime-M_{f_0}^2)
\nonumber
\\ &+& \rho^{cont}(s,s^\prime)\theta(s-s_0)
\theta(s^\prime-s_0^\prime)\,, \label{rho} \eea where
$\rho^{cont}$ includes the contribution of the higher resonances
and of the hadronic continuum. The resulting expression for $T_1$
is: \bea T_1(p^2,(p+q)^2)&=& {f_K M_{f_0} {\tilde f} g_{f_0 K^+
K^-} \over (M^2_K-p^2) (M_{f_0}^2-(p+q)^2)} \nonumber \\  &+&
\int_D ds ds^\prime {\rho^{cont}(s,s^\prime) \over (s-p^2)
[s^\prime-(p+q)^2]} \, . \label{disphad} \eea
We do not consider possible subtraction terms in eq.(\ref{disp})
as they will be removed by a Borel transformation.

For space-like and large external momenta (large $-p^2$,
$-(p+q)^2$)  $T_1$ can be computed in QCD as an expansion near the
light-cone $x^2=0$. The expansion involves matrix elements of
non-local quark-gluon operators,  defined in terms of kaon
distribution amplitudes of increasing twist. \footnote{The
short-distance expansion of the 3-point function of one scalar
$\bar s s$ and two pseudoscalar $\bar s i \gamma_5 q$ densities
was considered in \cite{Narison:1996fm}. The present calculation
mainly differs for the possibility of incorporating an infinite
series of local operators \cite{Colangelo:2000dp}.} The first few
terms in the expansion are retained, since the higher twist
contributions are suppressed by powers of  $1/(-p^2)$ or
$1/(-(p+q)^2)$. For the resulting expression for $T_1$, obtained
to twist four accuracy, we refer to \cite{Colangelo:2003jz}.

The sum rule for $g_{f_0 K^+ K^-}$ follows from the approximate
equality of eq.(\ref{disphad}) and the computation of $T_1$ in
QCD.  Invoking global quark-hadron duality, the contribution of
the continuum in (\ref{disphad}) can be identified with the QCD
contribution above the thresholds $s_0, s^\prime_0$. This allows
us to isolate the pole contribution in which the coupling appears.
Such a  matching  is improved performing two independent Borel
transformations with respect to the variables $-p^2$ and
$-(p+q)^2$, with  $M_1^2$, $M_2^2$ the Borel parameters associated
to the channels $p^2$ and $(p+q)^2$, respectively. In order to
identify the continuum contribution with the QCD term, a
prescription has been proposed in \cite{Belyaev:1994zk},
consisting in considering the symmetric values $M_1^2=M_2^2=2
M^2$.  Such a prescription is not adeguate in our case, where the
Borel parameters correspond to channels with different mass scales
and should not be constrained to be equal. A different method has
been suggested in \cite{Colangelo:2003jz} for the present
calculation, exploiting the property of the leading twist wave
functions of being polynomials in $u$ (or $1-u$). The subleading
twist terms represent a small contribution to the QCD side of the
sum rule, and hence the calculation can leave them unaffected.

\begin{figure}\vspace{-1cm}
  \resizebox{0.5\textwidth}{!}{\includegraphics{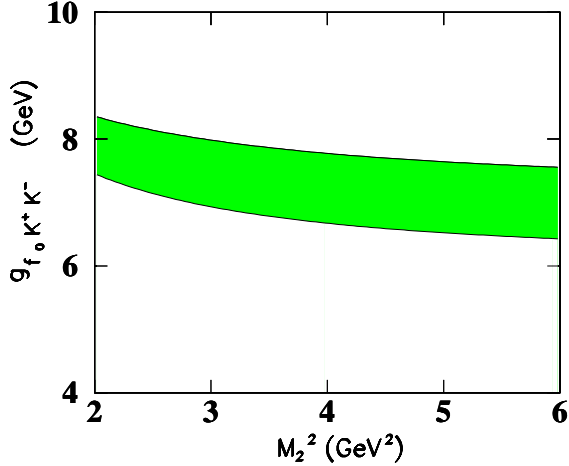}}\vspace{-1cm}
  \caption{$g_{f_0 K^+ K^-}$ as a function of the Borel
parameter $M_2^2$, varying: $1.05 \le s_0\le 1.15$ GeV$^2$ and
 $0.7 \le M_1^2 \le 2.0$ GeV$^2$.}
\label{fig:gf0kk}       
\end{figure}

The main nonperturbative input quantities  in the final sum rule
are the kaon light-cone wave functions. A theoretical framework
for their determination relies on an expansion in terms of matrix
elements of conformal operators \cite{Braun:1989iv}.  For the kaon
we took into account the meson mass corrections, related to the
parameter $\rho^2={m_s^2\over M_K^2}$, worked out in
\cite{Ball:1998je}.
 For details about the distribution amplitudes we refer to the
\cite{Belyaev:1994zk,Ball:1998je}. In the analysis of the sum rule
we use $m_s(1\,$GeV$)=0.14$ GeV \cite{ms}, $M_K=0.4937$ GeV,
$M_{f_0}=0.980$ GeV, $f_K=0.160$ GeV and ${\tilde f}=(0.180 \pm
0.015)$ GeV \cite{DeFazio:2001uc}. The threshold parameter $s_0$
is varied around the value $s_0=1.1$ GeV$^2$ fixed from the
determination  of $f_K$ using two-point sum rules \cite{fk}.
 The final sum rule provides $g_{f_0 K^+ K^-}$ as a function of
 the Borel parameters $M_1^2$, $M_2^2$.
A stability region where the outcome does not depend on $M_i^2$
can be selected. Such a region does not correspond to the line
$M_1^2=M_2^2$, but to the range $0.8 \le M_1^2 \le 1.6$ GeV$^2$
with $M_2^2$ extending up to $M_2^2\simeq 5$ GeV$^2$. Varying
$M_1^2$ and  $M_2^2$ in this region, and changing the values of
the thresholds and of the other parameters, we obtain the result
depicted in fig.\ref{fig:gf0kk}, which can be quoted as $6.2 \le
g_{f_0 K^+ K^-}\le 7.8$ GeV.

Let us briefly discuss the uncertainties affecting the numerical
result. We neglected the $SU(3)_F$ breaking effects which render
the kaon distribution amplitudes asymmetric with respect to the
middle point; such a  neglect should have a minor role in our
approach, as discussed in \cite{Colangelo:2003jz}. Another
uncertainty is related to the value of the strange quark mass,
$m_s$; since the dependence of the sum rule on $m_s$ mainly
involves the ratio $M_K^2/m_s$, one can fix this ratio using
chiral perturbation theory, obtaining results in the same range
quoted for $g_{f_0K^+K^-}$.

\section{Comparison with other results and conclusions}
\label{conclusions} The various  determinations of $g_{f_0K^+K^-}$
form a very complex
 scenario. A collection of experimental results is provided
in table \ref{tableexp}. In the case of KLOE Collaboration, two
results are reported, corresponding to two different fits
performed in the analysis of the data, indicated  with (A) and
(B). The difference mainly consists in the inclusion of the
$\sigma$ contribution in fit (B). Such a result is the  one
affected by the smallest uncertainty, and seems to point towards
large values of $g_{f_0K^+K^-}$. Theoretical results also  lie in
a rather large range of values, from $2$ GeV up to $7 $ GeV. For a
 detailed discussion  we refer to \cite{Colangelo:2003jz},
while an analysis based on experimental data can be found in
\cite{Boglione:2003xh}. The outcome of light-cone QCD sum rule
analysis, reported here, is in keeping with a large value for the
coupling. The uncertainty affecting the result is intrinsic of the
method and does not allow a better comparison with data. However,
the analysis confirms a peculiar aspect of the scalar states, i.e.
their large hadronic couplings, thus pointing towards a scenario
in which the process of hadronic dressing is favoured. However,
since  the most accurate experimental data stem from the
investigation of $\phi \to f_0 \gamma$,  it is mandatory to wait
for the study of unrelated processes, namely the combined analysis
of $D_s$ decays to pions and kaons, which could be performed, for
example, at the B-factories.

{\bf Acknowlwdgments} I thank P. Colangelo  for collaboration.
Partial support from the EC Contract No. HPRN-CT-2002-00311
(EURIDICE) is acknowledged.

\begin{table}
\caption{Experimental determinations of $g_{f_0 K^+ K^-}$ using
different physical processes. } \label{tableexp}
\begin{tabular}{llll}
\hline\noalign{\smallskip}
Collaboration &process & $g_{f_0 K^+ K^-}\,(GeV)$ & Ref. \\
\noalign{\smallskip}\hline\noalign{\smallskip} KLOE    &$\phi \to
f_0 \gamma \, (A)$&$4.0\pm 0.2 \, (A)$&
\cite{Aloisio:2002bt}\\
        &$\phi \to f_0 \gamma \, (B)$&$5.9\pm 0.1 \, (B)$&\\
CMD-2   &$\phi \to f_0 \gamma$       &$4.3\pm 0.5$       &
\cite{Akhmetshin:1999di} \\   SND     &$\phi \to f_0 \gamma$
&$5.6\pm 0.8$       & \cite{Achasov:2000ym} \\ WA102   &$pp$
&$2.2\pm 0.2$        & \cite{Barberis:1999cq} \\ E791    & $D_s\to
3 \pi$
&$0.5\pm 0.6$      & \cite{Gobel:2000es}\\
\noalign{\smallskip}\hline
\end{tabular}
\end{table}


\begin{thebibliography}{}
%

\bibitem{rassegna}
L. Montanet, Rep. Prog. Phys. \textbf{46}, (1983) 337;
T. Barnes, hep-ph/0001326.

\bibitem{Close:2002zu}
F.~E.~Close and N.~A.~Tornqvist,
J.\ Phys.\ G {\bf 28} (2002) R249.

\bibitem{Morningstar:1999rf}
C.~J.~Morningstar and M.~J.~Peardon, Phys. Rev. D \textbf{60},
(1999) 034509;
S.~Narison, Nucl. Phys. A \textbf{675}, (2000) 54C.

\bibitem{Close:1993ti}
F.~E.~Close {\it et al.}, Phys. Lett. B \textbf{319}, (1993) 291.

\bibitem{tornqvist}
N.A. Tornqvist, Phys. Rev. Lett. \textbf{49}, (1982) 624.


\bibitem{Tornqvist:1995kr}
N.A. Tornqvist,
Z. Phys. C \textbf{68}, (1995) 647.

\bibitem{roos}
N.A.Tornqvist, M.Roos,
Phys. Rev. Lett. \textbf{76}, (1996) 1575.

\bibitem{scadron}
E. van Beveren {\it et al.},
Z. Phys. C \textbf{30}, (1986) 615;
M.D. Scadron,
Phys. Rev.  D \textbf{26}, (1982) 239;
E. van Beveren {\it et al.},
Phys. Lett.  B \textbf{495}, (2000) 300.


\bibitem{jaffe}
R.L. Jaffe,
Phys. Rev.  D \textbf{15}, (1977) 267, 281; D \textbf{17}, (1978)
1444;
R.L. Jaffe, K. Johnson,
Phys. Lett. B \textbf{60}, (1976) 201.

\bibitem{ivan}
N.N. Achasov and V.N. Ivanchenko,
Nucl. Phys. B \textbf{315}, (1989) 465;
N.N. Achasov and V.V. Gubin,
Phys. Rev. D \textbf{56}, (1997) 4084.

\bibitem{isgur}
J. Weinstein and N. Isgur,
Phys. Rev. Lett. \textbf{48}, (1982) 659;
Phys. Rev. D \textbf{27}, (1983) 588;
Phys. Rev. D \textbf{41}, (1990) 2236.

\bibitem{closebook}
N. Brown and F.E. Close, \textit{The DA$\Phi$NE Physics HandBook}
(L. Maiani, G. Pancheri and N. Paver eds, INFN Frascati, 1995)
447; F.E. Close {\it et al.}, Nucl. Phys. B \textbf{389}, (1993)
513.

\bibitem{kaminski}
R. Kaminski {\it et al.},
Phys. Rev. D \textbf{50}, (1994) 3145.

\bibitem{shev}
N.N. Achasov {\it et al.},
Phys. Rev. D \textbf{56}, (1997) 203.

\bibitem{Tornqvist:2002bx}
N.~A.~Tornqvist,
arXiv:hep-ph/0204215.

\bibitem{Boglione:1996uz}
M.~Boglione and M.~R.~Pennington, Phys. Rev. Lett.  \textbf{79},
(1997) 1998.

\bibitem{Tornqvist:2000ju}
N.~A.~Tornqvist,
arXiv:hep-ph/0008136.

\bibitem{Shakin:gn}
C.~M.~Shakin, H.~Wang,
Phys. Rev. D \textbf{63}, (2001) 014019.

\bibitem{DeFazio:2001uc}
F.~De Fazio and M.~R.~Pennington,
Phys. Lett. B \textbf{521}, (2001) 15.


\bibitem{Close:2001ay}
F.~E.~Close and A.~Kirk, Phys. Lett. B \textbf{515}, (2001) 13.

\bibitem{Colangelo:2003jz}
P.~Colangelo, F.~De Fazio,
Phys. Lett. B \textbf{559}, (2003) 49.

\bibitem{lightconesr}
I.~I.~Balitsky {\it et al.},
Nucl. Phys. B \textbf{312}, (1989) 509;
V.~L.~Chernyak and I.~R.~Zhitnitsky,
Nucl. Phys. B \textbf{345}, (1990) 137.

\bibitem{Colangelo:2000dp}
For a recent review of the method see: P.~Colangelo and
A.~Khodjamirian, \textit{At the Frontier of Particle Physics, vol.
3} (M. Shifman ed., World Scientific, 2001)  1495.

\bibitem{Aliev:2001mm}
T.~M.~Aliev {\it et al.}, Phys. Lett. B \textbf{527}, (2002) 193.

\bibitem{Craigie:1982ng}
 N.~S.~Craigie and J.~Stern,
Nucl. Phys. B \textbf{216}, (1983) 209.

\bibitem{Narison:1996fm}
S.~Narison,
Nucl. Phys. B \textbf{509}, (1998) 312.

\bibitem{Belyaev:1994zk}
V.~M.~Belyaev {\it et al.},
Phys. Rev. D \textbf{51}, (1995) 6177.

\bibitem{Braun:1989iv}
V.~M.~Braun, I.~E.~Halperin,
Z. Phys. C \textbf{48}, (1990) 239.

\bibitem{Ball:1998je}
P.~Ball, JHEP \textbf{9901}, (1999) 010.

\bibitem{ms}
P.~Colangelo {\it et al.},
Phys. Lett. B \textbf{408}, (1997) 340;
K.~Maltman and J.~Kambor, Phys. Rev. D \textbf{65}, (2002) 074013;
M.~Jamin {\it et al.},
Eur. Phys. J. C \textbf{24}, (2002) 237.

\bibitem{fk}
A.A. Ovchinnikov and A.Pivovarov,  Phys. Lett. B \textbf{163},
(1985) 231.

\bibitem{Aloisio:2002bt}
A.~Aloisio {\it et al.},
Phys. Lett. B \textbf{537}, (2002) 21.

\bibitem{Akhmetshin:1999di}
R.~R.~Akhmetshin {\it et al.},
Phys. Lett. B \textbf{462}, (1999) 380;
Nucl. Phys. A \textbf{675}, (2000) 424C.

\bibitem{Achasov:2000ym}
M.~N.~Achasov {\it et al.},
Phys. Lett. B \textbf{485}, (2000) 349.

\bibitem{Barberis:1999cq}
D.~Barberis {\it et al.},
Phys. Lett. B \textbf{462}, (1999) 462.

\bibitem{Gobel:2000es}
E.~M.~Aitala {\it et al.},
Phys. Rev. Lett.  \textbf{86}, (2001) 765.


\bibitem{Boglione:2003xh}
M.~Boglione and M.~R.~Pennington,
arXiv:hep-ph/0303200.

\end{thebibliography}
\end{document}